\documentclass{article}
\usepackage{arxiv}

\usepackage{amsfonts}       %
\usepackage{subfigure}
\usepackage{caption}
\usepackage{mathtools}
\usepackage{amsmath,amsfonts,amssymb,amsthm}
\usepackage[numbers]{natbib}
\usepackage{doi}
\usepackage{commath}

\usepackage[utf8]{inputenc} %
\usepackage[T1]{fontenc}    %
\usepackage{hyperref}       %
\usepackage{url}            %
\usepackage{booktabs}       %
\usepackage{amsfonts}       %
\usepackage{nicefrac}       %
\usepackage{microtype}      %
\usepackage{lipsum}
\usepackage{graphicx}
\usepackage{algorithm}
\usepackage{algorithmicx}
\usepackage{algpseudocode}
\title{Finite Element Method-enhanced Neural Network for Forward and Inverse Problems}

\author{Rishith Ellath Meethal\\
	Technology\\
	Siemens AG\\
	Munich, Germany\\
	\texttt{rishith.ellath\_meethal@siemens.com} \\
	\And 
	Birgit Obst\\
	Technology\\
	Siemens AG\\
	Munich, Germany \\
	\texttt{birgit.obst@siemens.com} \\
	\And
	Mohamed Khalil\\
	Technology\\
	Siemens AG\\
	Munich, Germany\\
	\texttt{khalil.mohamed@siemens.com} \\
	\And
	Aditya Ghantasala\\
	Chair of Structural Analysis\\
	Technical University of Munich, \\
	Munich, Germany \\
	\texttt{aditya.ghantasala@tum.de} \\
	\And
	Anoop Kodakkal\\
	Chair of Structural Analysis\\
	Technical University of Munich, \\
	Munich, Germany \\
	\texttt{anoop.kodakkal@tum.de} \\
	\And
	Kai-Uwe Bletzinger\\
	Chair of Structural Analysis\\
	Technical University of Munich, \\
	Munich, Germany \\
	\texttt{kub@tum.de} \\
	\And
	Roland Wüchner\\
	Institute of Structural Analysis\\
	Technische Universität Braunschweig\\
	Braunschweig, Germany\\
	\texttt{r.wuechner@tu-braunschweig.de} \\
}

\begin{document}
\maketitle
\begin{abstract}
We introduce a novel hybrid methodology combining classical finite element methods (FEM) with neural networks to create a well-performing and generalizable surrogate model for forward and inverse problems. The residual from finite element methods and custom loss functions from neural networks are merged to form the algorithm. The Finite Element Method-enhanced Neural Network hybrid model (FEM-NN hybrid) is data-efficient and physics conforming. The proposed methodology can be used for surrogate models in real-time simulation, uncertainty quantification, and optimization in the case of forward problems. It can be used for updating the models in the case of inverse problems. The method is demonstrated with examples, and the accuracy of the results and performance is compared against the conventional way of network training and the classical finite element method. An application of the forward-solving algorithm is demonstrated for the uncertainty quantification of wind effects on a high-rise buildings. The inverse algorithm is demonstrated in the speed-dependent bearing coefficient identification of fluid bearings. The hybrid methodology of this kind will serve as a paradigm shift in the simulation methods currently used.
\end{abstract}

\keywords{
Hybrid models, Informed machine learning, Knowledge-based neural networks, Machine learning, FEM-based neural network, Self-supervised learning}

\section{Introduction}
Developments in the field of Artificial Intelligence (AI) \cite{russell2002artificial} have introduced a substantial improvement in everyday life. Advances in the sub-fields of AI such as Data science, Machine learning, Deep learning, Neural networks are contributing to diverse research and application fields. These include computer vision\cite{szeliski2010computer}, speech and language processing\cite{jurafsky2000speech}, drug discovery\cite{burbidge2001drug}\cite{lavecchia2015machine}, genomics\cite{libbrecht2015machine}, computer games \cite{funge2004artificial}, animation\cite{grzeszczuk1998neuroanimator}, robotics \cite{kawato1988hierarchical}, and many more. Deep learning has boosted the development of computer vision by contributing extensively to image classification \cite{lu2007survey}, object detection \cite{papageorgiou1998general}, and semantic segmentation \cite{long2015fully}. Similarly, computer games have also seen a large number of contributions \cite{nareyek2004ai} \cite{fairclough2001research} \cite{yannakakis2005ai} from AI. An important direction in this field is Game physics, where smoke and fluid flows for computer graphics are simulated with neural networks. Jonathan et al. \cite{DBLP:journals/corr/TompsonSSP16} proposed a data-driven solution to the inviscid-Euler equations, which is faster than traditional methods used in computer graphics animations. Similar to the developments in Game physics, different AI methods have found their application in solving Partial Differential Equations (PDE) for physical problems.

Recently, Neural networks\cite{hecht1992theory} are applied along with numerical methods for simulation \cite{thuerey2020deep} \cite{pfaff2020learning}. Conventional approaches typically use input and output data to create neural network-based surrogate models approximating the mapping function between them. Such surrogate models can be used for fast simulation. However, this approach has two significant shortcomings. 
The first one is the high computational cost of creating training data. This is because the simulation results required to make a surrogate model are generated by running large number of numerical simulations. Running a large number of simulations requires a large amount of computational power and simulation time. The second problem is that the training algorithm does not consider underlying physics. It results in neural networks that are loosely informed about the underlying physics. A loosely informed surrogate does not extrapolate well if the training data does not cover all the required range. However, it is crucial for a numerical simulation that the surrogate model created follows the underlying physics described by the PDE.

Recently there are developments in exploiting the prior information about the data by incorporating it into the learning process. Laura et al. \cite{von2019informed} provides a structured overview of various approaches to integrate different prior knowledge into the training process. The physics informed neural network (PINN) \cite{raissi2017physics} has shown how existing knowledge on the physics of the data can be used to constraint the neural network to follow the physics. This is accomplished by embedding the physics in the form of the partial differential equation into the custom loss of neural network using automatic differentiation \cite{griewank1989automatic}. This approach is used to solve both forward and inverse problems. It also shows that classical methods like Runge-Kutta time stepping can coexist with deep neural networks for time-dependent problems and offer better predictive algorithms. However, it is stated that the method does not replace classical numerical methods in terms of robustness and computational efficiency required in practice. 

In this contribution, we introduce a data-efficient and physics conforming hybrid methodology for generating numerically accurate surrogate models. It falls in the category of informed machine learning \cite{von2019informed}, where expert knowledge in the form of differential equations is used in the hypothesis set and learning algorithm of neural networks. 
The methodology involves training the neural network with FEM-based custom loss function and deploying it with the FEM. The use of FEM makes the training physics-conforming and the prediction error quantifiable. This algorithm is expanded for inverse problems as well. The rest of the paper is structured as follows: Section \ref{sec:algorithm} explains the algorithm used behind the novel hybrid model. Algorithms for both forward and inverse problems are explained here. The novelty brought in terms of data efficiency and physics conformity is compared with conventional neural network based surrogate models for simulation. The proposed model is demonstrated with examples in Section \ref{sec:examples}. The results are compared with FEM-based simulation and conventional neural networks. Applications of these methods for uncertainty quantification and parameter identification problems are discussed in Section \ref{sec:applications}. Section \ref{sec:conclusion} concludes the discussion. 

\section{Algorithm}
\label{sec:algorithm}

\subsection{Finite element method}
\label{sec:finite_element_method}
Numerical approximation of the continuous solution field $u$ of any partial differential equation (PDE) given by Equation \ref{Eq:ex_partial_opp} on a given domain $\Omega$ can be done using various methods. Among others, some of the widely used techniques are finite element method \cite{zienkiewicz2000finite}, finite volume method \cite{versteeg2007introduction}, particle methods \cite{onate2011particle}, and finite cell method \cite{kollmannsberger2019finite}. In this contribution, we restrict the discussion to Galerkin-based finite element methods.
\begin{align}
  \mathcal{L}(u)&=0 &\textnormal { on } &\Omega \label{Eq:ex_partial_opp}\\
  u&=u_d &\textnormal { on } &\Gamma_D \label{Eq:ex_partial_opp_bc_d}\\
  \frac{\partial u}{\partial x} &= g &\textnormal { on } &\Gamma_N \label{Eq:ex_partial_opp_bc_n}
\end{align}
Consider the PDE in Equation \ref{Eq:ex_partial_opp} defined on a domain $\Omega$ together with the boundary conditions given by Equations \ref{Eq:ex_partial_opp_bc_d} and \ref{Eq:ex_partial_opp_bc_n}. Here $u_d$ and $g$ are the Dirichlet and Neumann boundary conditions on the respective boundaries. A finite element formulation of Equation \ref{Eq:ex_partial_opp} on a discretization of the domain with $m$ elements and $n$ nodes, together with boundary conditions, will result in the system of Equations shown by Equation \ref{Eq:sys_linear_eq}. Here, we assume all the necessary conditions on the test and trial spaces \cite{zienkiewicz2000finite} are fulfilled.
\begin{equation}
  \underbrace{
    \begin{pmatrix}
     k_{1,1} & k_{1,2} & \cdots & k_{1,n} \\
     k_{2,1} & k_{2,2} & \cdots & k_{2,n} \\
     \vdots  & \vdots  & \ddots & \vdots  \\
     k_{n,1} & k_{n,2} & \cdots & k_{n,n}
    \end{pmatrix}
  }_{K(u^h)}
  \underbrace{
    \begin{pmatrix}
      u_1\\
      u_2\\
      \vdots\\
      u_n
    \end{pmatrix}
  }_{u^h}
    =
  \underbrace{
    \begin{pmatrix}
      F_1\\
      F_2\\
      \vdots\\
      F_n
    \end{pmatrix}
  }_F
  \label{Eq:sys_linear_eq}
\end{equation}
In Equation \ref{Eq:sys_linear_eq}, $K(u^h)$ is the non-linear left hand side matrix, also called the stiffness matrix. $u^h$ is the discrete solution field, and $F$ is the right hand side vector. The residual of the system of Equations in Equation \ref{Eq:sys_linear_eq} can be written as
\begin{equation}
  r(u^h) = K(u^h) u^h - F
  \label{Eq:residual_matix_eq_g}
\end{equation}
To obtain the solution $u^h$, a Newton-Raphson iteration technique can be employed using the linearization of $r(u^h)$ and its tangent matrix. This requires the solution of a linear system of equations in every iteration. These iterations are carried out until the residual norm $\norm{r}_n$ meets the tolerance requirements. For a detailed discussion of the methodology, the readers are referred to \cite{ypma1995historical}. For this residual based formulation, in case of a linear operator $\mathcal{L}$, it takes only one iteration to converge. For a large number of elements and nodes, among different steps of the finite element methodology, the most computationally demanding step is the solution of the linear system of equations. In an application where computational efficiency is critical, like real time simulations \cite{SurgerySimulationRealtime} and digital twin \cite{keiper2018reduced}, it is imperative that this step should be avoided. Techniques suitable for such applications, like model order reduction \cite{schilders2008model, ModelReductionAntoulas}, construct a surrogate model of Equation \ref{Eq:sys_linear_eq} to reduce this cost significantly. Techniques involving neural-networks, can completely avoid this cost, but will require a significant amount of training and test data, which is typically generated by simulation of the underlying finite element problem. 
In Section \ref{sec:fem_nn} we discuss an algorithm combining residual information from numerical method for the training of neural network for linear PDEs. In this case the residual $r(u^h)$ becomes 

\begin{equation}
r(u^h) = K u^h - F
\label{Eq:residual_matix_eq}
\end{equation}

\subsection{Neural network and custom loss}
\label{sec:nn_custom_loss}

Consider a neural network mapping from input variables $ \textbf{x} = [x_1, x_2,   ...  x_m]$ from domain $X$ to output variables $ \textbf{y} = [y_1,y_2, ... y_n]$ of domain $Y$. Input variables are given to the input layer of the neural network to get output variables from the output layer. The input layer is connected to the output layer by $L$ number of layers in between, called hidden layers. Each hidden layer receives input from previous layer and outputs $o^l = [o_1^l, o_2^l, ... ,o_k^l]$. The $w_l \in \mathbb{R}^{n_l \times n_l +1}$ matrix represents the weights between layer $l$ and $l+1$ and the vector $b_l$ of size $n_l$ represents the bias vector from layer $l$. Here $n_l$ represents number of neurons in each layer of the neural network. 

The output from any layer is transformed as below before sending it to the next layer. 
\begin{equation}
z^l = w^l  o^{l-1}+ b^l    
\end{equation}
 The nonlinear activation function $\sigma(.)$  is applied to each
component of the transformed vector $z^l$ before sending it as an input to the next layer.
\begin{equation}
o^l = \sigma(z^l)    
\end{equation}

Following this sequence for all the layers, the output of the entire neural network can be written as 
\begin{equation}
    y(\theta) = \sigma^L(z^L......\sigma^2(z^2.\sigma^1(z^1(x))))
\end{equation}

where $y$ is the output vector for the given input vector $x$, and $\theta = \{w^l, b^l\}_{l=1}^L $ is the set of trainable parameters of the network.

The network is trained by treating it as a minimization problem. The objective function for minimization is a function of $y$ and $y^t$, called as loss function among neural network community. Here $y$ is the predicted value from neural network and $y^t$ is the actual value which is either measured, or simulated.

The calculated loss $\delta$ is reduced by updating the trainable parameters in the process called backpropagation \cite{hecht1992theory}. In backpropagation, the gradient of $\delta$ with respect to trainable parameters is calculated, and the trainable parameters are updated as follows, 

\begin{equation}
    w_l = w_l - \eta \frac{\partial \delta}{\partial w_l}
\end{equation}

The derivatives of $\delta$ with respect to the trainable parameters are calculated using chain rule. The parameter $\eta$ is called learning rate and is chosen by the user. 

\begin{equation} 
	\begin{split}   	
	\frac{\partial \delta}{\partial w_l} &= \frac{\partial \delta}{\partial y} \frac{\partial y}{\partial w_l}                   \\ 	
    \frac{\partial \delta}{\partial b_l} &= \frac{\partial \delta}{\partial y} \frac{\partial y}{\partial b_l}                  
	\end{split}
	\label{equ:derivative_mse}
\end{equation}

In Equation \ref{equ:derivative_mse}, the derivative of output $y$ with respect to trainable parameters $\frac{\partial y}{\partial w_l} $ and $\frac{\partial y}{\partial b_l} $ are calculated by considering the networks architecture and activation functions. Implementations for the calculation of above mentioned derivatives are readily available in machine learning libraries like, TensorFlow \cite{abadi2016tensorflow}, PyTorch \cite {ketkar2017introduction}, etc. 
But the first part of Equation \ref{equ:derivative_mse} depends on the chosen loss function. One of the common loss functions for minimization among neural network communities is the mean square error (MSE) between the true value for $y^t$ and the predicted value for $y$. The true value is taken from the training data and the prediction is the network output. The loss function $\delta(y, y^t)$ is given by 
\begin{equation}
\delta_{MSE} = \dfrac{1}{m}{\sum}_{i=i}^m(y_i-y_i^t)^2
\end{equation}

For $\delta_{MSE}$ the first part of Equation \ref{equ:derivative_mse} is the derivative of $\delta_{MSE}$ with respect to output of the neural network $y$ given by, 
\begin{equation}
    \frac{\partial \delta}{\partial y} = 2\dfrac{1}{m}{\sum}_{i=i}^m(y_i-y_i^t)  
    \label{equ:derivative_mse_y}
\end{equation}

But there are large variety of neural networks which uses different other loss functions. Some statistical applications uses loss functions like mean absolute percentage error (MAPE) for training.

\begin{equation}
    \delta_{MAPE} = \dfrac{1}{m}{\sum}_{i=i}^m \left| \dfrac{(y_i-y_i^t)}{y_i^t} \right|
\end{equation}

In this case the first part of the Equation \ref{equ:derivative_mse}, $\frac{\partial \delta}{\partial y}$, will have a different function when compared with that used for MSE in Equation \ref{equ:derivative_mse_y}.

\subsection{Finite element method enhanced neural network for forward problems}
\label{sec:fem_nn}

\begin{figure}[ht]
    \centering
    \subfigure[FEM-neural network training]{\includegraphics[width=0.45\linewidth]{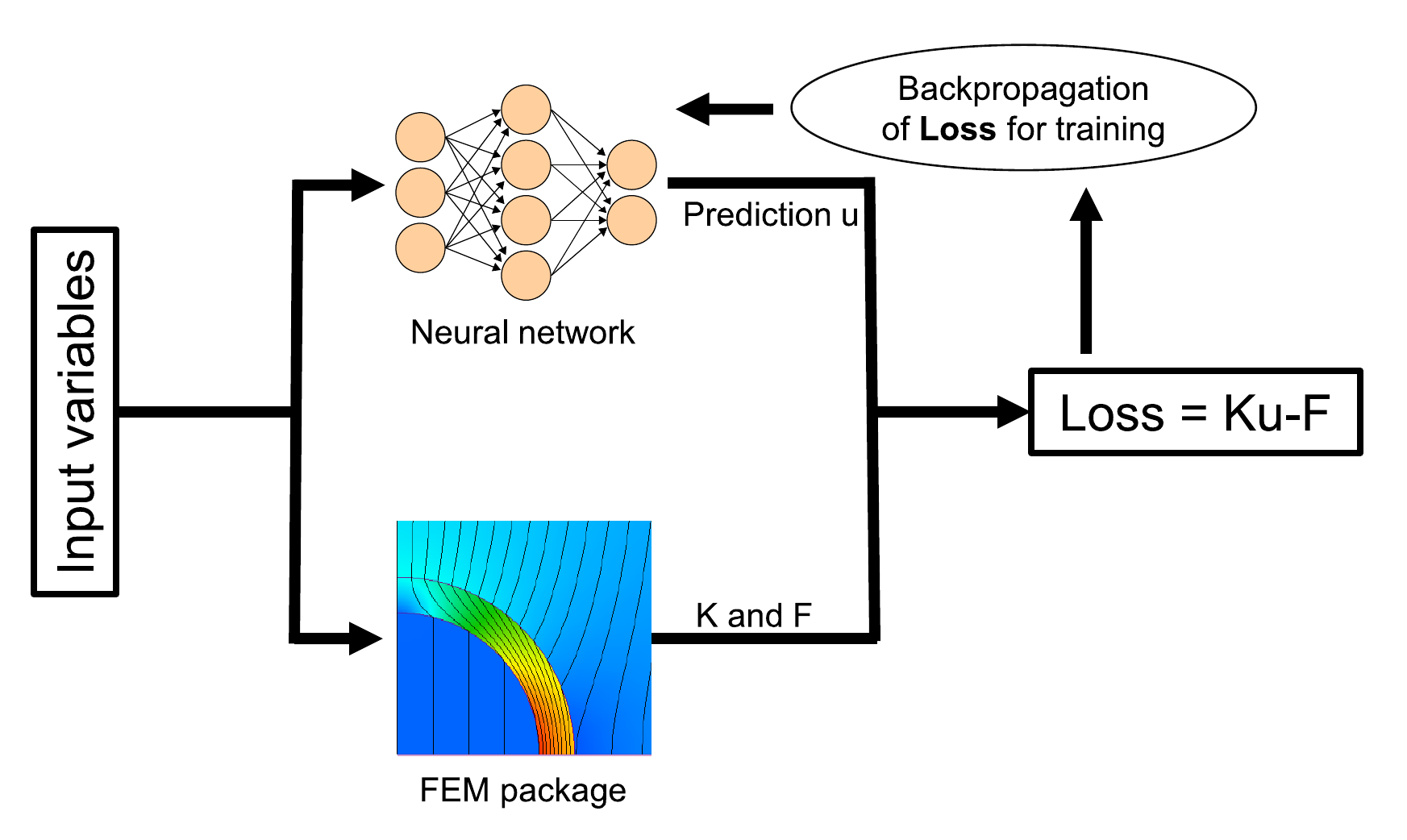} } 
    \subfigure[FEM-neural network deployment]{\includegraphics[width=0.45\linewidth]{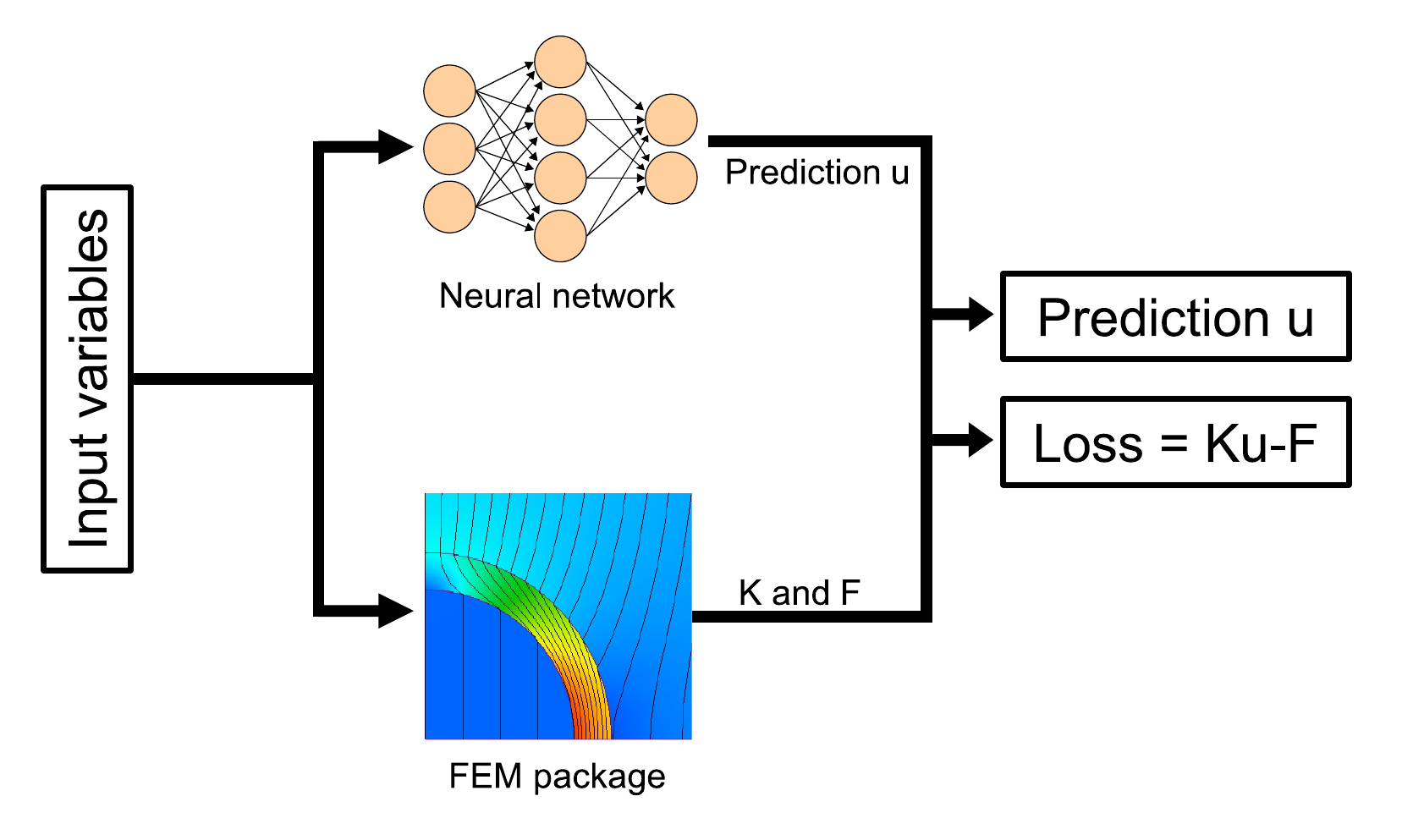}}
    \caption{FEM-NN algorithm for the training and deployment of the surrogate model}    
    \label{Fig:fem_nn_algorithm}
\end{figure}

The proposed algorithm is combining FEM and neural network to result in a surrogate model, that we call as FEM enhanced neural network hybrid model. The model undergoes training process before deploying it as a surrogate model for simulation. During the training process, the variables for simulation are taken as input parameters. The input parameters are processed by both the neural network and the FEM library. The Neural network outputs $u$ after the forward pass through the chosen network architecture. The output of the network is the discrete solution field vector $u$ given as, 

\begin{equation}
    U = \begin{pmatrix}
    u_1 \\
    u_2 \\
    \vdots \\
    u_n 
    \end{pmatrix} 
\end{equation}

In the mean time the FEM library takes the input variables and calculates the stiffness matrix $K$ and the force vector $F$ as explained in Section \ref{sec:finite_element_method}.

The residual $R$ is calculated using the prediction $U$ from the neural network and $K$ and $F$ from FEM. Loss for the neural network prediction is defined as Euclidean norm of the residual vector $r$. 
\begin{equation}
	\delta = {\vert \vert r \vert \vert}_2  
\end{equation}

where $r$ is given by 
\begin{equation}
\begin{split}
    r &=  KU-F \\
    &= 
\begin{pmatrix}
     k_{1,1} & k_{1,2} & \cdots & k_{1,n} \\
     k_{2,1} & k_{2,2} & \cdots & k_{2,n} \\
     \vdots  & \vdots  & \ddots & \vdots  \\
     k_{n,1} & k_{n,2} & \cdots & k_{n,n}
    \end{pmatrix}
    \begin{pmatrix}
      u_1\\
      u_2\\
      \vdots\\
      u_n
    \end{pmatrix}
    -  
    \begin{pmatrix}
      F_1\\
      F_2\\
      \vdots\\
      F_n
    \end{pmatrix}\\
    &= \begin{pmatrix}
    k_{1,1}u_1+k_{1,2}u_2+..+k_{1,n}u_n-F_1  \\
    k_{2,1}u_1+k_{2,2}u_2+..+k_{2,n}u_n-F_2  \\
    \vdots \\ 
    k_{n,1}u_1+k_{n,2}u_2+..+k_{n,n}u_n-F_n  \\
    \end{pmatrix} 
\end{split}
\label{eq:residual_fem}
\end{equation}
This gives loss $\delta$ as, 
\begin{equation}
\begin{split}
    \delta =  {\Vert r \Vert}_2  &= \sqrt{(k_{1,1}u_1+k_{1,2}u_2+..+k_{1,n}u_n-F_1)^2+...+(k_{n,1}u_1+k_{n,2}u_2+..+k_{n,n}u_n-F_n)^2} \\ 
 &= \sqrt{{\sum}_{j=1}^n {\sum}_{i=1}^n  (K_{j,i}u_i- F_j)^2}
\end{split}
\label{eq:custom_loss_fem}
\end{equation}

As explained in Section \ref{sec:nn_custom_loss}, the learnable parameters are updated using backpropagation to minimize the loss. Backpropagation calculates the gradients of the loss with respect to the learnable parameters. Since we use a loss function specific to FEM, we need to calculate the second part of the Equation \ref{equ:derivative_mse}, $ \frac{\partial \delta}{\partial y}$, for the custom loss used here. In the case of a hybrid model, the output $y$ of the neural network is the predicted discrete solution field $u$.

\[ y = \begin{pmatrix} u_1 \\ u_2 \\ \vdots \\ u_n \end{pmatrix} \]
The gradient has to be calculated with respect to each member of the output. $\frac{\partial \delta}{\partial y}$ becomes,
\begin{equation*}
\begin{split}  
    \begin{pmatrix}
    \frac{\partial \delta}{\partial u_1 } \\
    \frac{\partial \delta}{\partial u_2 } \\
    \vdots \\
    \frac{\partial \delta}{\partial u_n }
    \end{pmatrix}
    &= 
    \frac{1}{2\delta} 2       
    \begin{pmatrix}
    k_{1,1}u_1+k_{1,2}u_2+..+k_{1,n}u_n-F_1 \\
    \vdots \hspace{5mm} \\
    k_{n,1}u_1+k_{n,2}u_2+..+k_{n,n}u_n-F_n
    \end{pmatrix}^T
    \begin{pmatrix}
     k_{1,1} & k_{1,2} & \cdots & k_{1,n} \\
     k_{2,1} & k_{2,2} & \cdots & k_{2,n} \\
     \vdots  & \vdots  & \ddots & \vdots  \\
     k_{n,1} & k_{n,2} & \cdots & k_{n,n}
    \end{pmatrix} 
    \end{split}
\end{equation*}
which gives 
\begin{equation}
	\frac{\partial \delta}{\partial y} = \frac{r^T K }{\delta}   
\end{equation}
where $r^T$ is the transpose of the residual vector $r$ and $K$ is the stiffness matrix. With the implementation of the above in the machine learning frameworks, we can train the network against the residual of the differential equation. 
Second part of the Equation \ref{equ:derivative_mse} is readily available in all the neural network frameworks. 

\begin{algorithm} [ht]
\begin{algorithmic}[1]
\Procedure{TRAIN}{}
\State Read simulation parameters
\State Initialize neural network
\State Initialize weights and biases  
\State Let $L$ be the number of layers in the neural network
\State Initialize FEM Package 
\State Compute the system matrices $K$ and $F $
\While {not Stop Criterion } 

\State $u \leftarrow$ Neural network prediction
\State Compute the residual $r = K U - F$
\State Compute the loss as the Euclidean norm of the residual vector $\delta = ||r||_2$
\State Compute the derivative $\frac{\partial \delta }{\partial u} = \frac{r^T \times  K }{\delta}$ 
\ForAll {$l \in \{1, \dots, L\}$}
\State Compute the derivative of neural network parameters using chain rule $\frac{\partial \delta }{\partial w_l} = \frac{\partial \delta }{\partial u} \frac{\partial u }{\partial w_l} $
\State Update trainable parameters (weights and biases) $w_l = w_l - \eta  \frac{\partial\delta}{\partial w_l}$
\EndFor
\EndWhile
\EndProcedure
\vspace{1em}
\Procedure{PREDICT}{}
\State Initialize FEM Package 
\State Compute the system matrices $K$ and $F$
\State Predict $U$ using the trained neural network
\State Compute the Residual $r = K U - F$
\State \textbf{return} $U $ and $ r$
\EndProcedure
\end{algorithmic} 
\caption{FEM-NN training and prediction for forward problems}
\label{alg:algorithm1}
\end{algorithm}

The procedure for training and prediction is detailed in Algorithm \ref{alg:algorithm1}. Once trained, the neural network can be deployed using a similar hybrid approach. During the deployment of the network, the input variables pass through both the neural network and the FEM framework. The trained neural network predicts the output $u$. In a conventional way of deployment of networks, the prediction accuracy is not measurable. But here the output is used along with the stiffness matrix $K$ and force vector $F$ from FEM to calculate the residual $r$ in Equation \ref{Eq:residual_matix_eq} . The residual $r$ is a measure on how much the output deviates from the actual solution of the governing equation. Hence, the accuracy of prediction of FEM-neural network is measurable.

\subsection{Finite element method enhanced neural network for inverse problems}
\label{sec:fem_nn_inverse}

Inverse problems appear in many different applications of science and industry. Forward problems estimate the results for a defined cause, whereas inverse problems typically estimate the cause for the observed results. It is possible to use forward-solving software for inverse analysis as well. In such cases, the inverse problem is formulated as a parameter identification problem, where the unknown parameters of the forward problem are calculated by minimizing a suitable cost function. The estimation of unknown parameters results in correcting or updating the mathematical model used. Such a corrected or updated model can be used for different applications like simulation and prediction, optimization, system monitoring, fault detection, etc. This makes inverse problems very vital in engineering. The algorithm introduced in Section \ref{sec:fem_nn} is extended for inverse problems as well. 

In the case of inverse problems for Equation \ref{eq:residual_fem}, the primary variable $u$ is known whereas forces $F$ or stiffness matrix $K$ can have unknown parts. As there are different kinds of inverse problems, we consider one category where stiffness matrix have unknown parts for the rest of the paper. Such a problem can be described using the following equation

\begin{equation}
	\begin{pmatrix}
	K_{k} & K_{ku} \\
	K_{ku} & K_{u} 
	\end{pmatrix}
	\begin{pmatrix}
	 U_{k} \\
	 U_{u} 
	\end{pmatrix}
	= 
	\begin{pmatrix}
	F_{k} \\
	F_{u} 
	\end{pmatrix}
\end{equation}

Where $K_k$ is the known part and $K_u$ the unknown part of the system matrix. The $K_{ku}$ and $K_{uk}$ are the contribution of bearing DOFs to the remaining DOFs of the system matrix. They are zero unless there is any physical connection to the bearings. Similarly, $U_u$ are the responses at the bearing and $U_k$ are the responses at rest of the rotor. The loss for the neural network prediction is defined as Euclidean norm of the residual vector $r$. 
\begin{equation}
\delta = {\vert \vert r \vert \vert}_2  
\end{equation}

where $r$ is given by 
\begin{equation}
\begin{split}
r &=  KU-f \\
&= 
\begin{pmatrix}
K_{k} & K_{ku}  \\
K_{uk} & K_{u} \\
\end{pmatrix}
\begin{pmatrix}
U_k\\
U_u\\
\end{pmatrix}
-  
\begin{pmatrix}
F_k\\
F_u\\
\end{pmatrix}\\
&= \begin{pmatrix}
K_{k}U_k+ K_{ku}U_u-F_k  \\
K_{uk}U_k+ K_{u}U_u-F_u  \\
\end{pmatrix} 
\end{split}
\label{eq:residual_fem_inverse}
\end{equation}
This gives loss $\delta$ as, 
\begin{equation}
\delta =  {\left|\left|r \right|\right|}_2  = \sqrt{(K_{k}U_k+ K_{ku}U_u-F_k)^2+ (K_{uk}U_k+ K_{u}U_u-F_u)^2} \\ 
\label{eq:custom_loss_fem_inverse}
\end{equation}

Similar to the calculation done for forward problems, we need to calculate the derivative of residual with respect to neural network prediction to perform the backpropagation. In the case of Equation \ref{eq:custom_loss_fem_inverse} it is $\frac{\partial \delta}{\partial K_u}$

\begin{equation}
\frac{\partial \delta}{\partial K_u} 
= 
\frac{1}{\delta}     
(K_{uk}U_k+ K_{u}U_u-F_u) U_u
\label{eq:gradient_inverse_alg}
\end{equation}

Equation \ref{eq:gradient_inverse_alg} along with the second part of the Equation \ref{equ:derivative_mse} are used to update the neural network parameters for training the neural network to identify the unknown part of the matrix. 

\begin{algorithm} 
\begin{algorithmic}[1]
\Procedure{TRAIN}{}
\State Read simulation parameters
\State Initialize neural network
\State Initialize weights and biases
\State Initialize FEM Package 
\State Let $L$ be the number of layers in the neural network
\While {not Stop Criterion}
\State Compute the system matrices  $ K_k, K_{ku}, K_{uk},  F_k, F_k, U_k$ and $U_u $
\State $K_u \leftarrow$ Neural network prediction
\State Compute the residual $r = K U - f$
\State Compute the loss as the Euclidean norm of the residual vector $\delta = ||r||_2$
\State Compute the derivative $\frac{\partial \delta }{\partial K_u} = \frac{1}{\delta}(K_{uk}U_k+ K_{u}U_u-F_u) U_u $ 
\ForAll {$l \in \{1, \dots, L\}$}
\State Compute the derivative using chain rule $\frac{\partial \delta }{\partial w_l} = \frac{\partial \delta }{\partial K_u}  \frac{\partial K_u }{\partial w_l}  $
\State Update trainable parameters (weights and biases) $w_l = w_l - \eta  \frac{\partial\delta}{\partial w_l}$
\EndFor
\EndWhile
\EndProcedure
\vspace{1em}
\Procedure{PREDICT}{}
\State Initialize FEM Package 
\State Predict $K_u$
\State Use the predicted $K_u$ for forward simulation or other analysis
\EndProcedure

\end{algorithmic} 
\caption{FEM-NN training for inverse problems}
\label{alg:algorithm2}
\end{algorithm}

Procedure for training and predicting for an inverse problem is detailed in \ref{alg:algorithm2}. One example for such an inverse problems is the stiffness identification of fluid bearings in a rotor-dynamic system demonstrated in Section \ref{sec:fluid_bearing}.

\subsection{Comparison between FEM enhanced neural network and conventional neural networks for simulation} 
\label{sec:femnn_compare}

In this section the discussed novel FEM-NN hybrid model for simulation and the conventional way of making neural network-based surrogates for simulation are compared. 

Conventionally, a large number of simulations are run for making the input and output data required for the training of the neural network. Hence the training is a supervised training requiring a large amount of data. There are two main problems associated with this approach when it comes to numerical simulation of physical phenomena. First, such models are constructed without considering the physics or the differential equation behind the problem. During the training phase, the network treats the input and output of the simulation as pure data. The equations governing the simulation or the physics behind the problem are not considered. The second one is that a large amount of samples are required for training the network. Since the neural network training requires a large number of input-output pairs, we need to run a large number of simulations. Running such a large number of simulations is computationally expensive, especially when it comes to multiphysics simulations. In a nutshell, followings are the shortcomings associated with traditional neural network surrogates. 
 
 \begin{itemize}
  \item Neural network treats inputs and outputs as pure numbers without treating them as physical quantities.
   \item Training is done against the output data, not against the governing PDE describing the physics.
   \item Prediction can go wrong in an untrained scenario and the error of prediction is not measurable.
   \item A large number of computationally expensive simulations are required to create data. 
\end{itemize}

The proposed FEM NN solves the problems associated with conventional surrogates. The proposed algorithm uses the custom loss function defined in Equation \ref{eq:custom_loss_fem}. The custom loss is based on the discretized version of the PDE, hence it follows the physics. Since the loss is based on the predictions from neural network and matrices from FEM, it does not need target values or precalculated simulation results. Hence the computationally expensive simulations need not be run. Though there are methods like PINNs, they fail to converge to actual solution due to multiple reasons. One of the main reason is the multiobjective treatment of boundary conditions and PDE. This problem is overcome by using the introduced FEM-NN method.

\begin{itemize}
\item The network is trained against the equation, hence the physics is preserved.
\item The computational cost of simulation is less as the training does not require target values 
\item The prediction comes with the residual of the linear system in Equation \ref{Eq:residual_matix_eq}. This makes the prediction accuracy measurable. 
\item Single loss term taking care of boundary conditions and conservation laws

\end{itemize}

\section{Examples}
\label{sec:examples}

\subsection{Steady state convection diffusion problem}
\label{subsec:conv_diff_example}
This section discusses the application of FEM-based neural network applied to the one-dimensional convection-diffusion equation. The equation describes physical phenomena where 
particles, energy, or other physical quantities are transferred inside a physical system due to two processes; convection and diffusion. In this example, we consider the temperature as the physical quantity.      

\begin{figure}[t]
    \centering
    \subfigure[]{\includegraphics[width=0.45\textwidth]{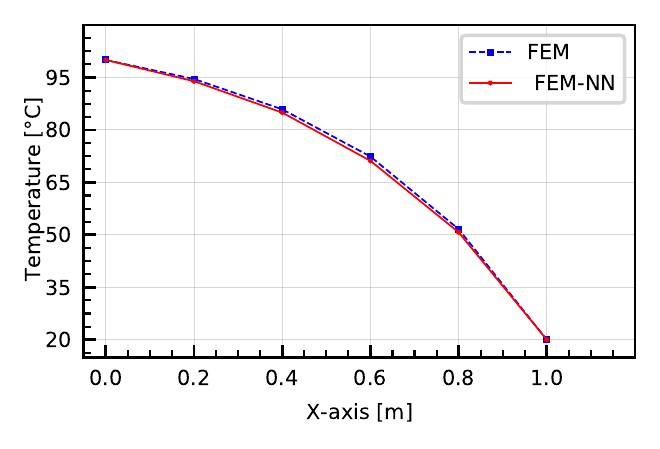} } 
    \subfigure[]{\includegraphics[width=0.45\textwidth]{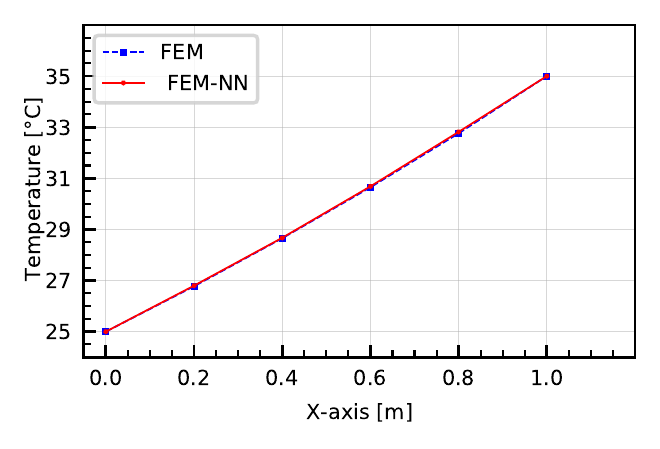}}
    \subfigure[]		  {\includegraphics[width=0.45\textwidth]{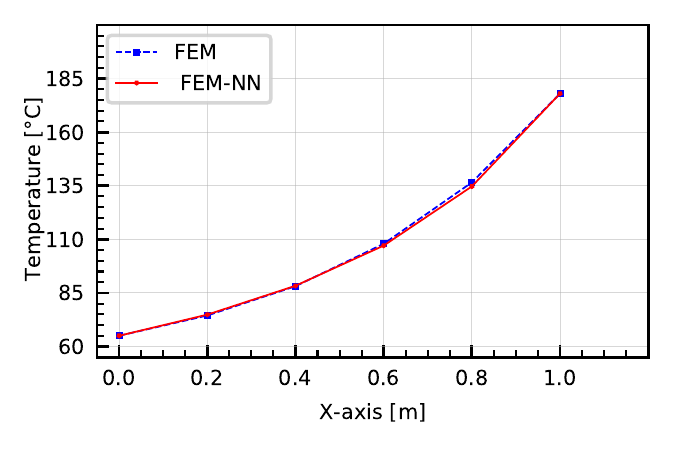}}  
    \subfigure[]{\includegraphics[width=0.45\textwidth]{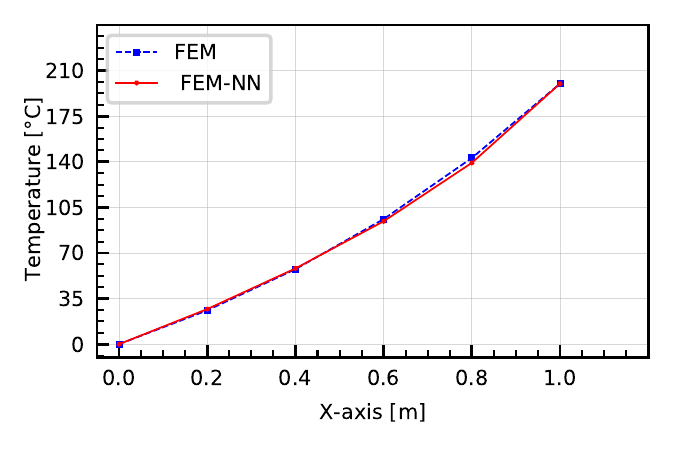}}
    \caption{Distribution of temperature along x-axis for the steady state convection diffusion example for cases (a) $T1 = 100, T2 = 20, k = 10, u = 20, S = 100$ (b) $T1 = 25, T2 = 35, k = 10, u = 3, S = 1$ (c) $T1 = 65, T2 = 178, k = 6, u = 11$ and (d) $T1 = 0, T2 = 200, k = 10, u = 30$}    
    \label{Fig:result_conv_diff}
\end{figure}

One-dimensional steady state convection diffusion equation for temperature is of the form
\begin{equation}
    \begin{multlined}
        \frac{\partial T}{\partial t} + u \frac{\partial T}{\partial x} = k \frac{\partial^2 T}{\partial x^2}  + S \:, 
    \label{equ:conv_diff}
    \end{multlined}
\end{equation} 
 
\[T(x=0) = T_1,  \quad  T(x=1) = T_2 \]

where $T$ is the temperature, $u$ the convection velocity, $k$ the thermal diffusion coefficient and $S(x)$ the heat source. The term $\frac{\partial T}{\partial t} = 0$ for steady state problems. The $2^{nd}$ order ODE is supported by two boundary conditions (BCs) provided at the two ends of the 1D domain, given as $T_1$ and $T_2$. We are concerned with the temperature distribution along the x-axis for the given boundary conditions, velocity, diffusion coefficient and source.

The model is trained using boundary conditions $T_1$ and $T_2$, source $S$, thermal diffusivity $k$ and convection velocity $u$ as the input parameters for getting the nodal temperature values as output. The parameters for the neural network $T(x)$ is learned by minimizing the FEM-based loss. Standard Adam optimizer \cite{kingma2014adam} is applied to minimize the following loss function,

\begin{equation}
    \delta = \frac{1}{N} {\sum}_{i=i}^N  {\vert \vert K_iT_i-F_i \vert \vert}_2^2
\label{eq:femloss_t}
\end{equation}

where $K$ is the stiffness matrix for the given input, $F$ the force vector and $T$ the output from the neural network and $N$ is the total number of samples used for training. Equation \ref{eq:femloss_t} guarantees the model learns and preserves the underlying physics than mere data.

The output of the model for different values of $u$, $k$ and $S$ under the boundary conditions $T_1$ and $T_2$ is given in Figure \ref{Fig:result_conv_diff}. The figure also includes standard FEM result for comparison. It can be observed that the model learns the physics, and predictions match with the results obtained using FEM. Figure \ref{Fig:result_conv_diff} has four sub-figures, which show the prediction for different combinations of $k$, $u$ and $S$. In every combination of $k$, $u$, $S$, $T1$ and $T2$, the network was able to predict the temperature distribution that closely matches with the FEM result. The average absolute error between FEM results and FEM-NN results are 0.616, 0.022, 0.352, and 0.642 degrees for Figures \ref{Fig:result_conv_diff}a, \ref{Fig:result_conv_diff}b, \ref{Fig:result_conv_diff}c and \ref{Fig:result_conv_diff}d respectively. 

The Figures \ref{Fig:result_conv_diff}a and \ref{Fig:result_conv_diff}b have constant heat source on every node. The Figures \ref{Fig:result_conv_diff}c and \ref{Fig:result_conv_diff}d have different values of heat source on each node. It also uses random values for other parameters too. The parameters used are, $T_1 = 65 $, $T_2 = 178$, $S = [5, 2, 3, 4, 5, 1 ]$, $u = 11$, $k = 6$ for Figure \ref{Fig:result_conv_diff}c. Similarly the parameters for the Figure \ref{Fig:result_conv_diff}d are $T_1 = 0 $, $T_2 = 200$, $S = [5, 2, 3, 4, 5, 1 ]$, $k = 1$, and $u = 1$. It can be observed that model generalizes quite well in predicting the distribution of temperature.

\begin{figure}[ht]
     \centering
        \includegraphics[width=.6\linewidth]{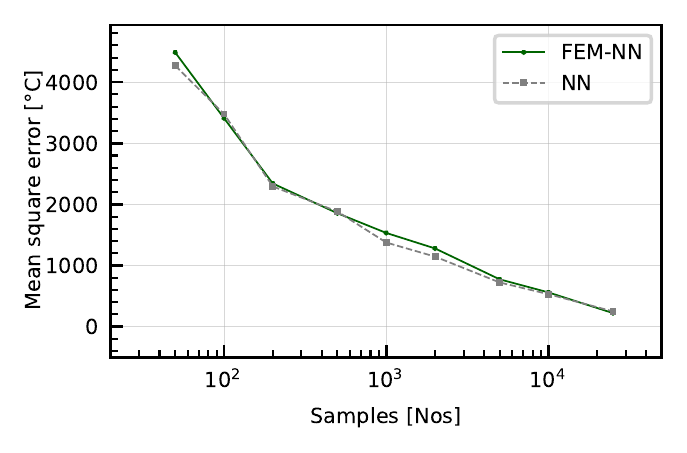}
     \caption{Comparison of error of prediction between FEM-NN and conventional neural network} \label{Fig:comparison_femnn_nn_conv_diff}
\end{figure}

This algorithm, as described in Section \ref{sec:fem_nn} does not use target values since it uses the matrix and vector from FEM for loss calculation. Hence, it does not fall in supervised learning and can be regarded as a semi-supervised learning approach. A comparison of the accuracy of prediction for FEM-NN and conventional neural networks is presented in Figure \ref{Fig:comparison_femnn_nn_conv_diff}. The loss is the $L_2$ of the difference between the prediction and the actual solution on 50000 samples. The actual solution of 50000 samples is created using a standard FEM code. 
It can be observed that the error is less or similar to that of the conventional neural network, whereas the time taken for conventional neural network training, including data creation, is 3.4 times more than the introduced FEM-NN.

\subsection{23 member truss}
\label{subsec:truss_example}

\begin{figure}[ht]
    \centering
    \includegraphics[width=\textwidth]{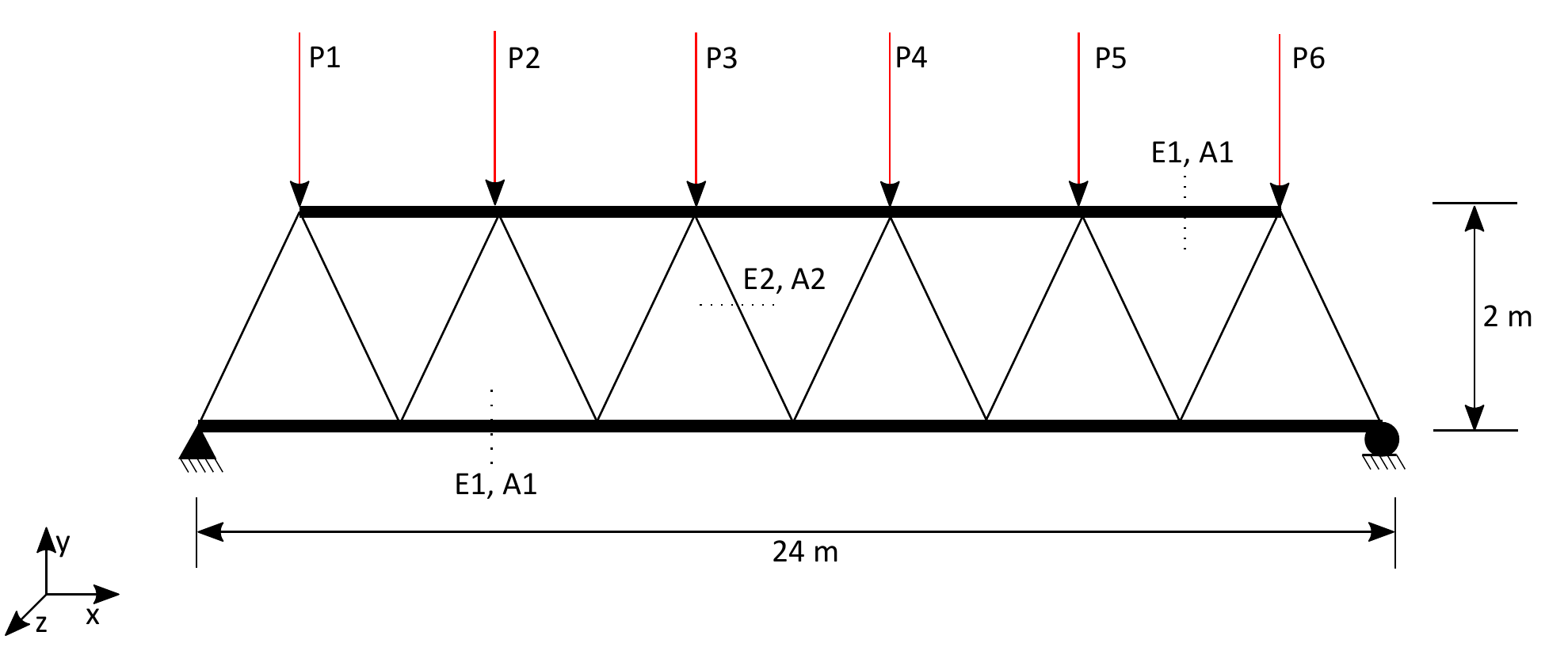}
    \caption{23 member truss with load}    
    \label{Fig:sudret_truss}
\end{figure}

In this example we consider a 23 member simply supported truss structure example taken from \cite{sudret2007uncertainty}. The geometrical dimensions of the truss structure are given in Figure \ref{Fig:sudret_truss}. The Young's modulus of horizontal bars are given by $E_1$ and vertical are given by $E_2$. The horizontal bars have a cross sectional area of $A_1$ and vertical bars have $A_2$. The truss is loaded by vertical forces $P1-P6$. All these 10 variables are taken as input to the neural network to predict the 39 nodal displacements for the 13 nodes of the truss. The mean and standard deviation of the training sample for each variable used as input is given in Table \ref{tab:input_sudret}.

\begin{table}[ht]
\caption{Input parameters for the 23 bar truss problem}
\begin{center}
\begin{tabular}{lccc}
\toprule
\textbf{Input variable}  &\textbf{Distribution}  & \textbf{Mean}  & \textbf{Standard deviation} \\
\midrule
Horizontal cross section area $A_h (m^2)$ & Normal & $1.0 \times 10 ^{-3} $  & $1.0 \times 10 ^{-4} $ \\
Vertical cross section area  $A_v (m^2)$ & Normal & $2.0 \times 10 ^{-3} $  & $2.0 \times 10 ^{-4} $ \\
Horizontal Young's modulus $E_h (Pa)$ & Normal & $2.1 \times 10 ^{11} $  & $2.1 \times 10 ^{10} $ \\
Vertical Young's modulus  $E_v (Pa)$ & Normal & $2.1 \times 10 ^{11} $  & $2.1 \times 10 ^{10} $ \\
 Vertical forces $P1-P6 (N)$ & Normal & $-5.0 \times 10 ^{5} $  & $5.0 \times 10 ^{4} $ \\
\bottomrule
\end{tabular}
\end{center}
\label{tab:input_sudret}
\end{table}

\begin{figure}[ht]
    \centering
    \includegraphics[width=\textwidth]{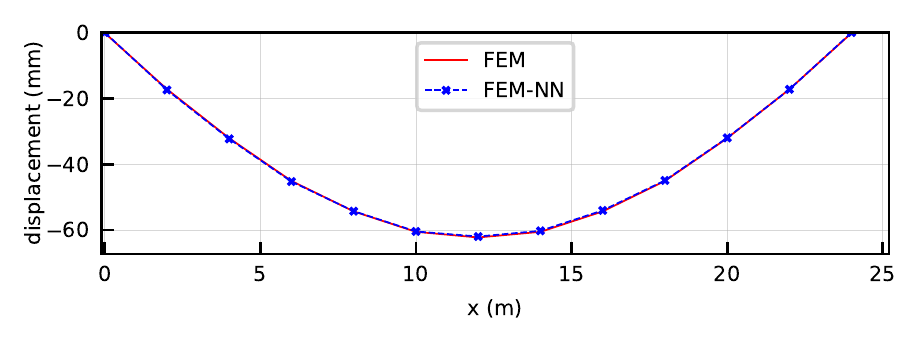}
    \caption{Vertical deflection of the truss at lower horizontal section}    
    \label{Fig:sudret_truss_deflection}
\end{figure}

The vertical displacement of the nodes using FEM-NN surrogate model is plotted in Figure \ref{Fig:sudret_truss_deflection}. The example shown took the inputs $E_1=2.1\times 10 ^{11}$, $E_2=2.32\times 10 ^{11}$, $A_1=9.2\times 10 ^{-4}$, $A_2=1.89\times 10 ^{-3}$, $P_1=-5.2\times 10 ^{4}$, $P_2=-5.2\times 10 ^{4}$, $P_3=-5.4\times 10 ^{4}$, $P_4=-3.6\times 10 ^{4}$, $P_5=-6.5\times 10 ^{4}$ and $P_6=-4.4\times 10 ^{4}$. The displacement is compared with a reference solution calculated using standard FEM. It can be observed that the surrogate model predicts the displacements accurately. The mean error in prediction is as low as $1e^{-4}$. 

\section{Applications}
\label{sec:applications}
\subsection{Uncertainty quantification of vibration due to wind load on high-rise building }
\label{subsec:wind_example}
It is important to identify, characterize and calculate the uncertainties on the results of analyses of complex systems. It is mandatory to have uncertainty analysis as per many regulatory standards and guidelines. Oberkampf et al. \cite{oberkampf2002error} says, “realistic, modeling and simulation of complex systems must include the non-deterministic features of the system and environment”.

Sampling-based uncertainty analysis with Monte Carlo approaches is widely used in the characterization and quantification of uncertainty \cite{helton1994treatment}. Monte Carlo based uncertainty analysis can now be found in virtually all engineering fields \cite{janssen2013monte}. Such analysis leads to final results expressed as a complementary cumulative distribution function (CDF). A large number of simulations, each with different input values sampled from their respective distributions, are run for Monte Carlo based uncertainty analysis. The results from simulations are used to obtain probability distributions of targeted outcomes. Hence, the Monte Carlo is inherently computationally expensive. Two methods are mainly used to reduce the number of simulation runs. The first one is improving the efficiency of sampling strategy while attaining the desired accuracy with a minimum number of simulation runs. The second one is stopping the analysis once suitably accurate results have been reached. 
But it is also possible to use surrogate models with less computational time instead of the original simulation model. A trained neural network can be used as a surrogate model. This way, the overall cost of Monte Carlo uncertainty analysis can be reduced. One major drawback of using surrogate models is their reduced accuracy. Since the conventional neural networks behave as a black box, we can not quantify the accuracy of the prediction. 

\begin{figure}[b]
     \centering
        \includegraphics[width= 0.5\linewidth, ]{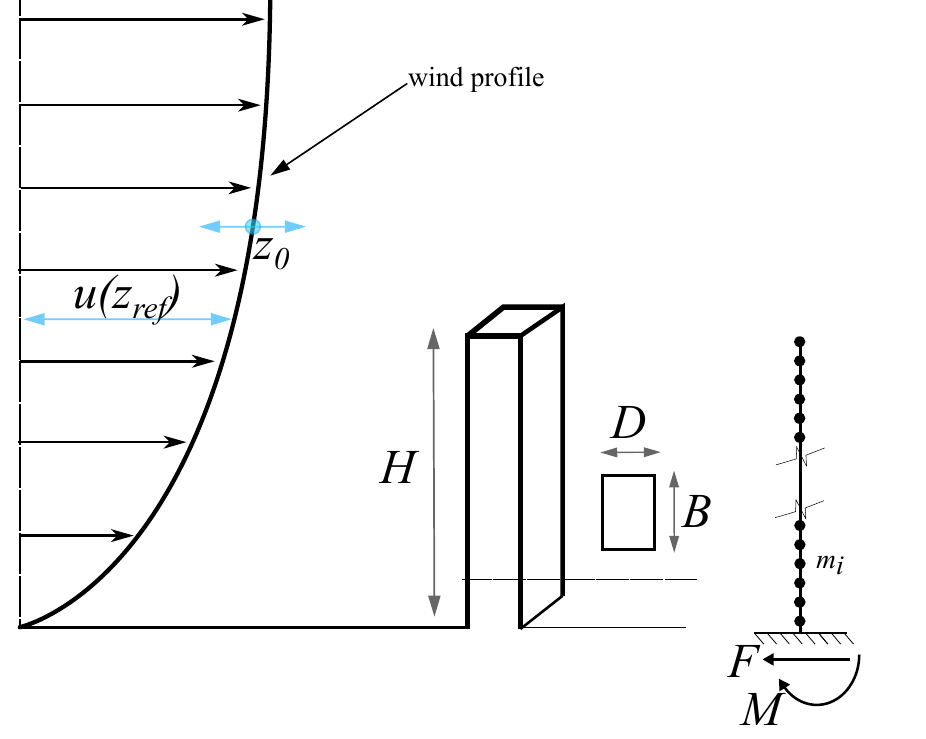}
     \caption{Wind load on a high-rise building} \label{Fig:wind_load}
\end{figure} 

The FEM-NN can be used as an alternative to overcome such challenges. The integral nature of using FEM along with neural network with residual-based training enables us to get the residual of each prediction. This way, some of the predictions which are not accurate enough can be solved further using any of the conventional iterative method, using the prediction from FEM NN as the initial value. Wind load on a high-rise building is used to demonstrate the use of FEM-NN for uncertainty quantification analysis.

\begin{table}[t]
\parbox{.49\linewidth}{
\centering
\caption{Details of the building - geometry and structural}
\begin{tabular}{lr}
\toprule
\textbf{Parameters}                  & \textbf{Values}             \\
\midrule
Height (H)                           & 180 m                        \\ 
Width (W)                            & 45 m                         \\ 
Length (D)                           & 30 m                         \\ 
Frequency (f)                        & 0.2 Hz                      \\ 
Density ($\rho$)        & 160 kg/m$^3$ \\ %
Damping ratio ($\zeta$) & 0.01                        \\
\bottomrule
\end{tabular}
\label{tab:highrise}
}
\hfill
\parbox{.49\linewidth}{
\centering
\caption{Details of uncertain wind parameters}
\begin{tabular}{lcr}
\toprule
\multicolumn{1}{l}{\textbf{\begin{tabular}[c]{@{}c@{}}Uncertain \\ parameters\end{tabular}}} & \multicolumn{1}{c}{\textbf{Distribution}} & \multicolumn{1}{r}{\textbf{Values}}       \vspace{0.3em}                                  \\ \midrule
\begin{tabular}[c]{@{}l@{}}Mean wind \\ velocity $u(z_{ref})$\end{tabular}                     & Weibull                                    & \begin{tabular}[r]{@{}r@{}}Mean = 40 m/s\\ Shape parameter = 2\end{tabular} \vspace{1em} \\
\begin{tabular}[c]{@{}l@{}}Roughness \\ length, $z_0$\end{tabular}                             & Uniform                                    & {[}0.1, 0.7{]}                                                               \\ \bottomrule
\end{tabular}
\label{tab:highrise_2}
}

\end{table}

The wind load on high-rise buildings is studied during the design of buildings as it can cause structural damage to the buildings. The uncertainty in wind climate and uncertain terrain also affect the long term performance of the structure. Hence the uncertainty in these parameters and its effects are to be studied. The effect of uncertainty in the wind load on the horizontal displacement at the top of the building is studied in this example. The CAARC \cite{braun2009aerodynamic} building geometry is used for the study. The parameters of the building height (H), width (B) and length (D), air density $\rho$, natural frequency of the building (f) and damping coefficient ($\zeta$) are shown in Figure \ref{Fig:wind_load} and given in Table \ref{tab:highrise}. The high-rise building is modeled using Euler-Bernoulli beam model as described by the following equations
\begin{equation}
        \frac{\partial^2 }{\partial x^2}( EI \frac{\partial^2 u_z(x)}{\partial x^2} ) = f (x)
    \label{equ:euler_ber_x}
\end{equation}

\begin{equation}
        \frac{\partial^2 }{\partial x^2}( EI \frac{\partial^2 u_y(x)}{\partial x^2} ) = f (x)
    \label{equ:euler_ber_y}
\end{equation} 

Where $u_y$ and $u_z$ represents the deflection of the building and $f$ represents the applied load. $E$ represents the Young's modulus and $I$ is the second moment of area. The corresponding surrogate model is created using FEM-NN. The wind profile and surface roughness are taken as input to the model. The mean and standard deviation of the training sample for each variable used as input is given in Table \ref{tab:highrise_2}. A wind-load on the building is calculated using the mean wind velocity $u(z_{ref})$ and roughness length $z_0$.  The static wind load at each height of the building is calculated as 
\begin{equation}
	F_d(z) = \frac{\rho {V(z)}^2 A C_d }{2}
\end{equation}
where, $\rho$ is the air density, $V(z)$ is the velocity at height $z$, $A$ is the reference area and $cd$ is the coefficient of drag for the cross section.

\begin{figure}[ht]
    \centering
    \subfigure[]{\includegraphics[width=0.45\textwidth]{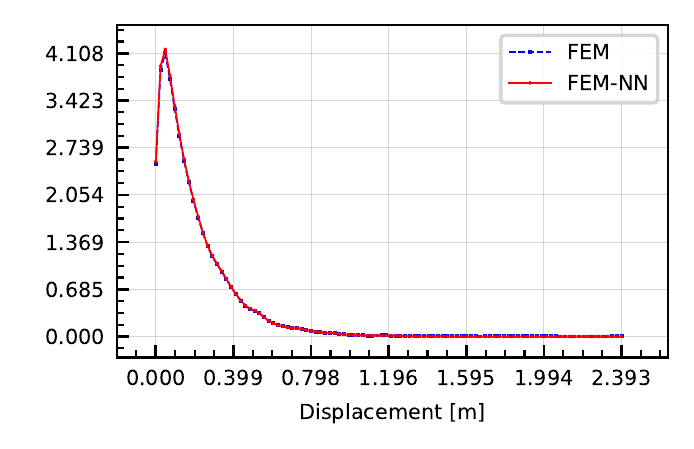}}
    \subfigure[]{\includegraphics[width=0.45\textwidth]{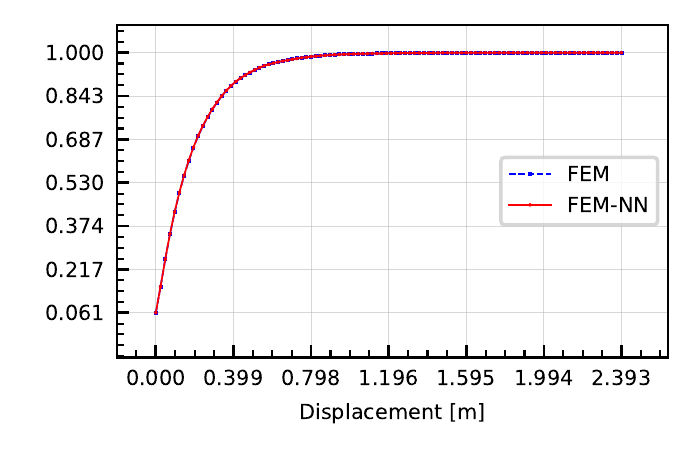}}
    \subfigure[]{\includegraphics[width=0.45\textwidth]{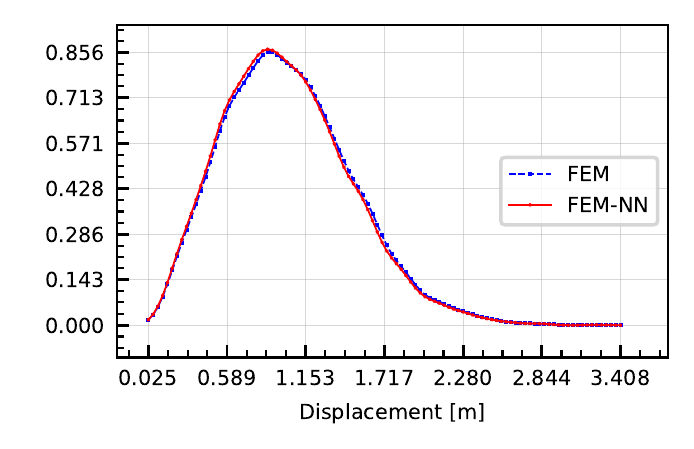}}  
    \subfigure[]{\includegraphics[width=0.45\textwidth]{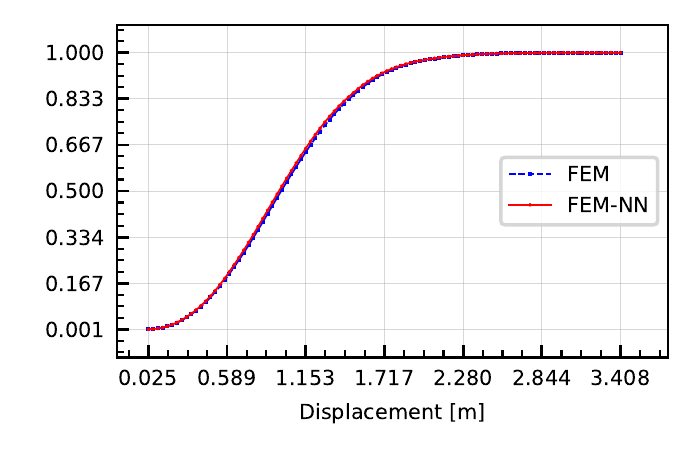}}
    \caption{PDF and CDF in trained region in (a) and (b) and untrained region in (c) and (d)}  
    \label{Fig:femnn1_uq}
\end{figure}

The trained model is used for quantifying the uncertainty associated with the effects of wind load on a high-rise building. The trained model is used to run a large number of simulations to perform the uncertainty quantification using Monte-Carlo method. 

Uncertainty in calculating the top displacement of the building is plotted for different cases in Figure \ref{Fig:femnn1_uq}. The Figure \ref{Fig:femnn1_uq}~(a) shows the probability distribution of the top displacement of the building. The Figure \ref{Fig:femnn1_uq}~(b) shows the cumulative distribution function. Both figures also contain the results obtained using FEM-based simulations. It can be observed that the FEM-NN is able to reproduce the results obtained using FEM with reasonable accuracy. In this case, the input parameters were taken from the same distribution as that used for training. The Figures \ref{Fig:femnn1_uq} (c) and (d) are the results of analysis from a different distribution for input parameters than that used for training. The statistical quantities of the analysis for both the cases are given in Table \ref{table:stat_quantities}. The FEM-NN was able to produce accurate results in this case too. Hence, the FEM-NN can replace the conventionally expensive FEM model for analysis like Monte-Carlo uncertainty quantification. 

\begin{table}[b]
\caption{Statistical quantities of Monte-Carlo analysis }
\centering
\resizebox{\textwidth}{!}{%
\begin{tabular}{|l|l|l|l|l|l|l|l|l|}
\hline
 & \multicolumn{4}{c|}{{ Trained region}} & \multicolumn{4}{c|}{Untrained region} \\ \hline
 & mean & \begin{tabular}[c]{@{}l@{}}standard \\ deviation\end{tabular} & skewness & kurtosis & mean & \begin{tabular}[c]{@{}l@{}}standard \\ deviation\end{tabular} & skewness & kurtosis \\ \hline
FEM & 0.183 & 0.199 & 2.679 & 12.278 & 1.016 & 0.470 & 0.667 & 0.717 \\ \hline
FEM-NN & 0.181 & 0.197 & 2.706 & 12.555 & 1.004 & 0.468 & 0.693 & 0.790 \\ \hline
\end{tabular}%
}
\label{table:stat_quantities}
\end{table}

\begin{figure}[!h]
     \centering
        \includegraphics[width=.6\linewidth]{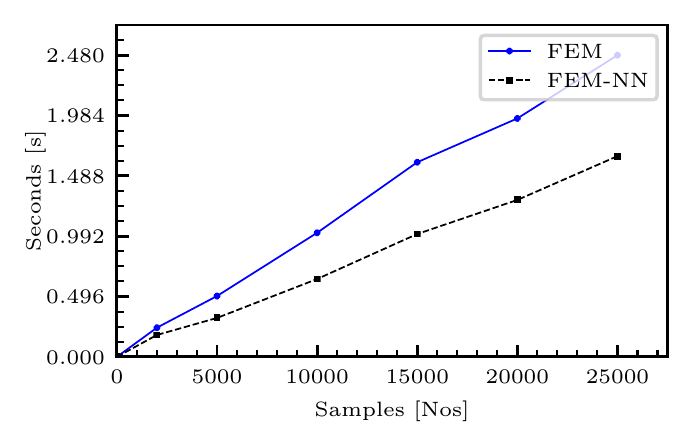}
     \caption{Cost for Monte-Carlo simulations in seconds} \label{Fig:cost_MC}
\end{figure}

The cost for running Monte-Carlo based uncertainty quantification using FEM and FEM-NN is plotted against different number of samples in Figure \ref{Fig:cost_MC}. It can be seen that the FEM-NN based analysis reduces the cost significantly. 

\subsection{Fluid bearing stiffness identification}
\label{sec:fluid_bearing}
Rotor dynamic systems are of engineering interest because of their usage in a wide range of applications like power plants, engines, etc. The accurate prediction of its dynamic behavior is vital for uninterrupted operation and safety. The estimation of bearing coefficients has been a primary barrier for predicting or simulating the dynamic behavior of such systems. For example, the dynamic coefficients of fluid bearings vary with the rotating speed of the machine. Fluid bearings are bearings in which the load is supported by a thin layer of fluid. In hydro-static bearings, the fluid is pressurized using an external pump, whereas, in hydrodynamic bearings, the fluid is pressurized with the speed of the rotating shaft/journal. A detailed review of different parameter identification methods for rotor dynamic systems can be found in \cite{lees2004identification}. 

The introduced FEM-NN can be used as an alternative to the different existing methods for bearing parameter identification. In this example, the stiffness coefficients of fluid bearings for the rotor dynamic system are modeled as a neural network taking rotational speed as input.

\begin{figure}[ht]
	\centering
	\includegraphics[trim={3cm 7cm 19cm 6cm}, clip, width=0.5\linewidth]{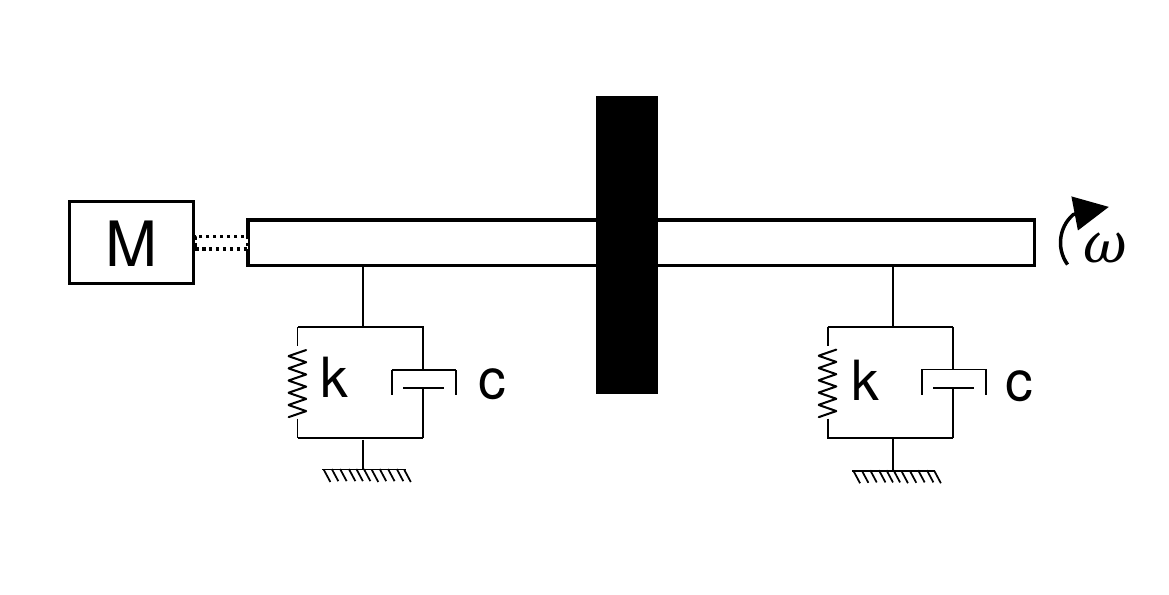}
	\caption{Schematic diagram of a flexible rotor-bearings system supported on bearings} \label{Fig:rds_sp_damper}
\end{figure}

\begin{equation}
	K_b = f (\omega)
	\label{eq:st_nn}
\end{equation}
where $K_b$ represents the dynamic stiffness coefficients of the fluid bearing and $\omega$ represents the rotational speed of the shaft. The function $f$ represents the neural network mapping rotational speed to stiffness coefficients. The complete system of equations describing the motion of a rotor-dynamic system in frequency domain can be described using the equation \ref{eq:rds}

\begin{equation}
\left[ -\omega^2 M+j{\omega}(G{\omega}+C)+K \right] q = F 
\label{eq:rds}
\end{equation}

Where $M$, $G$, $C$ and $K$ are mass, gyroscopic, damping and stiffness matrices for the complete system. $F$ and $q$ are the force and response of the system in the frequency domain. When the stiffness of fluid bearing is modeled as Equation \ref{eq:st_nn}, the system stiffness matrix $K$ can be written as 
\begin{equation}
	K = K_r + K_b
\end{equation}
where $K_r$ is the stiffness matrix of the system excluding bearings. 

\begin{figure}[ht]
	\centering
    \includegraphics[width=\textwidth]{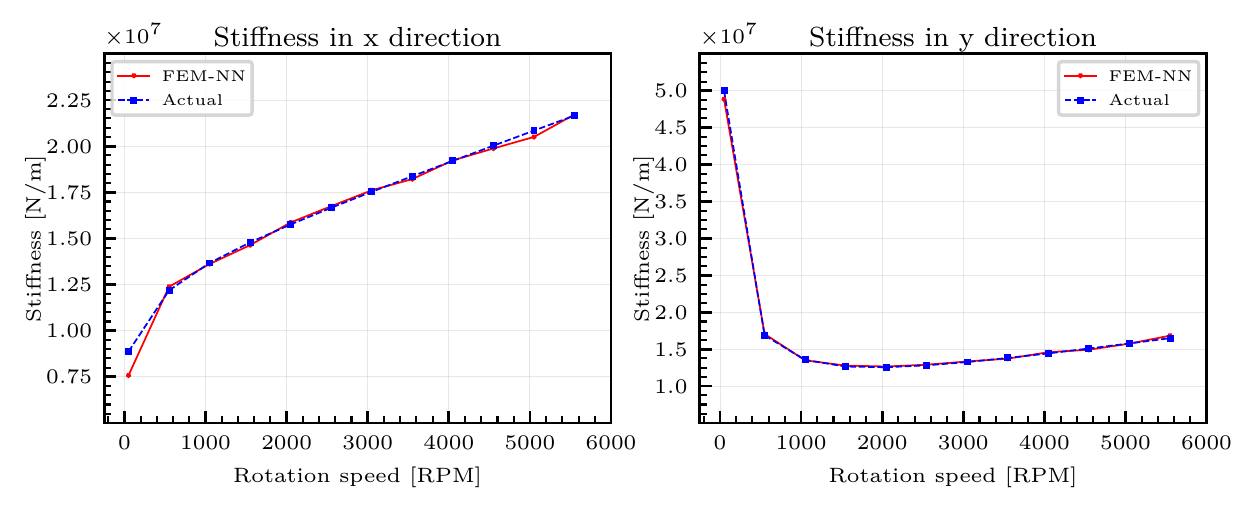} 

	\caption{Bearing stiffness vs Speed for x-direction (left) and y-direction (right)} 
	\label{Fig:bearing_stiffness}
\end{figure}

Similar to the forward problems, inverse problems algorithm also uses the residual of the equation as the loss function. In this case the residual $r$ is 
\begin{equation}
r = \left[ -\omega^2 M+j{\omega}(G{\omega}+C)+K \right] q - F 
\end{equation}
Minimizing this residual by optimizing the neural network parameters results in a neural network which predicts the bearing stiffness for a given speed.

The Figure \ref{Fig:bearing_stiffness} compares the predicted bearing stiffness against the actual. The actual stiffness values are calculated by modelling a fluid bearing with the help of an open source software called ross rotordnamic \cite{ross2020}. It can be observed that the stiffness variation against speed of rotation is captured precisely by the neural network. The stiffness in x and y direction, which vary differently, are learned by a single training process using the FEM-NN algorithm for inverse problems proposed in \ref{sec:fem_nn_inverse}.

\section{Conclusion and outlook} 
\label{sec:conclusion}
The proposed method suggests coupling the expertise from numerical methods and artificial intelligence and benefits from each.
It has multiple advantages compared to any data-driven model used so far in the literature. Instead of following the typical supervised learning approach, this algorithm uses a semi-supervised model. Integration of the underlying PDE equations in the discretized form into the custom loss function of the neural network enables the model to conform to the physics behind the problem rather than blindly learning from the data. 
This method saves the computation cost of running the simulations to get input-output pairs for supervised learning. Instead, only the inputs and an evaluation of the physics-based loss function is needed for the model training.
Additionally, developments in both the FEM community and the AI community could drive further development of the hybrid approach. 

Though the algorithm is simple and has the potential to change the simulation methods drastically, it demands more research. The present method developed for linear problems needs to be investigated for non-linear cases. The constant parameterized input nature of neural networks restricts using a trained neural network on other geometries. Though concepts like transfer learning offer flexibility, it requires further investigation to be combined with the proposed algorithm. The same problem can be solved from the numerical community point of view by employing a polycube representation of the geometry. However, such methods are yet to be explored.

\bibliography{mybibfile}

\end{document}